\documentclass[prl,twocolumn,amsmath,amssymb,aps,a4page]{revtex4}
\usepackage{graphicx}
\usepackage{amsmath,amssymb,mathrsfs}

\def\Tr{\operatorname{Tr}} 
\def\>{\rangle}\def\<{\langle} \def\sH{\mathscr{H}}
   
\def\mE{\mathcal{E}} \def\aE{\widetilde{\mathcal{E}}}
\def\dual#1{#1^\prime} 
 \def\fid{\mathscr{F}}
\def\N#1{|\!|{#1}|\!|} \def\Fcorr{F_{e,\textrm{a}}^{\mM}(\rho)}
\def\mutinf{I_{S:E}^{\mM}} \def\mM{\mathbf{M}}
\def\supp{\mathsf{Supp}}

\begin{document}

\title{Channel correction via quantum erasure}

\author{Francesco Buscemi} \affiliation{ERATO-SORST Quantum
  Computation and Information Project, Japan Science and Technology
  Agency,\\ Daini Hongo White Building 201, 5-28-3 Hongo, Bunkyo-ku,
  Tokyo 113-0033, Japan.} \email{buscemi@qci.jst.go.jp}

\date{\today}

\begin{abstract}
  By exploiting a generalization of recent results on
  environment-assisted channel correction, we show that, whenever a
  quantum system undergoes a channel realized as an interaction with a
  probe, the more efficiently the information about the input state
  can be erased from the probe, the higher is the corresponding
  entanglement fidelity of the corrected channel, and vice-versa. The
  present analysis applies also to channels for which perfect quantum
  erasure is impossible, thus extending the original quantum eraser
  arrangement, and naturally embodies a general
  information-disturbance tradeoff.
\end{abstract}

\maketitle

In quantum cryptography, the ability of faithfully transmit arbitrary
quantum states between two parties is a sufficient condition to
achieve privacy with respect to a third malicious party, which is
usually described as the purifying environment. The converse statement
is not true, in the sense that it has been recently proved that it is
possible to distill private states from channels which have zero
capacity~\cite{zerocap}. There are common situations, however, in
which the third party is not malicious, but (in some degree) helpful.
This is the case of \emph{environment-assisted channel
  correction}~\cite{GW}. This Letter shows that, in this case, privacy
and quantum capacity are equivalent quantities: the assisting party
can choose to give up (to \emph{erase}) some of her information about
the transmitted states in order to help the other two parties to
enhance the capacity of their quantum communication, and, the more she
erases, the better is the corresponding assisted quantum capacity.

A paramount \emph{ante litteram} example of assisted channel
correction is the \emph{quantum eraser}~\cite{eraser}, that is, a
variety of the usual double-slit interference experiment with a single
particle, in which it is possible to mark either particle- or
wave-like (exploring also halfway~\cite{englert-prl}) properties of
the beam by measuring one observable among a set of non-commuting
observables of the probe, which, previous to the measurement, has been
made suitably interact with the beam in order to store the
\emph{which-path} information.  Such which-path information sits in
the correlations established during the interaction between the
particle and the probe. In particular, Ref.~\cite{EB00} extends such
kind of duality relations to more general situations, in which a
two-level system interacts with a probe in such a way that
measurements on the latter result in a sorting of the former into
sub-ensembles exhibiting particle- or wave-like characteristics. It
holds that, the more which-path information the measurement collects
from the probe, the less visible are the fringes in the conditional
sub-ensembles, and vice-versa.

Another analogous situation can be recognized in the
\emph{partial teleportation} of an unknown quantum state. In the ideal
scenario~\cite{tele}, the two parties share a maximally entangled
state, and the assisting party performs a Bell measurement jointly on
the unknown state and on her branch of the shared entangled resource,
telling the result to the assisted party. In this case, the assisting
party is left with no information about the input state, while the
assisted party can perfectly recover it with unit probability. The
imperfect situation~\cite{partele} happens when the shared state is
not maximally entangled, or when the measurement is not a complete
Bell measurement.  In this case, the teleportation is noisy, in the
sense that it happens either probabilistically or with a fidelity
smaller than one. Correspondingly, the assisting party is left with
\emph{some} information about the state to be
teleported~\cite{partele}. Also in this case one would say, closely
mirroring the quantum eraser situation, that, the less information the
assisting party collects about the unknown state, the better is the
quality of the teleported state at the assisted party's side, and
vice-versa.

The two examples of quantum eraser and partial teleportation suggest
that, whenever a quantum system interacts with an assisting
environment (or probe), it may be possible to restore coherence lost
under the effect of a noisy channel by ``erasing'' from the
probe---that is, by performing on the probe a measurement whose
outcomes are as much independent as possible of---the information
carried by the input system itself. In the present Letter we will show
that it is indeed possible to formalize such an insight and to extend
the mentioned approach on a general basis. More explicitly, we will
consider general quantum evolutions, mathematically described as
channels---i.~e. completely positive trace-preserving
maps~\cite{kraus}---acting on an input system and physically modeled
as unitary interactions of the input system with a probe, the latter
playing the role of a controllable environment.  Then, by exploiting
the theory of environment-assisted correction~\cite{GW}, we will
derive two inequalities relating the amount of information (about the
input state) extracted from the probe with the entanglement fidelity
of the corresponding corrected channel, showing that perfect
erasure---that is, null information---is equivalent to perfect
correction and, even if perfect erasure is impossible, a robust
tradeoff relation between information extraction and channel
correction efficacy holds, and an optimal correction scheme can be
explicitly constructed. In this sense we can think that a sort of
\emph{quantum erasure relation} holds valid in all conceivable
situations, also providing, as a byproduct, a quite general
information-disturbance tradeoff relation.

\emph{Environment-assisted channel correction.---} Let us given a
channel $\mE$ acting on density matrices $\rho$ defined on the (finite
dimensional) input Hilbert space $\sH_S$. Basically, there exist two
equivalent ways to represent a channel, both highly non unique. The
first one is the Kraus representation~\cite{kraus}, that is,
\begin{equation}\label{eq:kraus}
  \mE(\rho)=\sum_kE_k\rho E_k^\dag,
\end{equation}
where the operators $\{E_k\}_k$ satisfy normalization
$\sum_kE_k^\dag E_k=\openone$. The second one, coming as a direct
consequence of the Stinespring theorem~\cite{stine}, regards the
channel as the average of an indirect measurement scheme, in which
first the system interacts with an environment (described by the
Hilbert space $\sH_E$ and initialized in a pure state $|0\>_E$), and
then a Positive-Operator--Valued Measure (POVM) $\mM=\{M_j^E\}_j$,
$M_j^E\ge 0$ $\forall j$, $\sum_jM_j^E=\openone$, is measured on the
environment, in formula~\cite{ozawa}
\begin{equation}\label{eq:unitary-real}
  \mE(\rho)=\sum_j\Tr_E[U(\rho\otimes|0\>\<0|_E)U^\dag\ (\openone\otimes M_j^E)].
\end{equation}
(In order to keep the notation simple, we consider, without loss of
generality, the output systems $S$ and $E$ to be equal to the input
ones.) The task is to correct the channel $\mE$. Without having access
to the environment (and hence to the assisting indices $j$), the best
one can do is described in Ref.~\cite{schum-west}, in which it is
shown that, if the coherent information is close to the input entropy,
it is possible to devise a correcting channel $\mathcal{C}^\rho$,
depending in general also on the input state $\rho$, such that
$\mathcal{C}^\rho\circ\mE$ is close to the identity channel on the
support of $\rho$.

The assisted scenario, first introduced in Ref.~\cite{GW}, is much
more powerful, since we now allow the environment to be somehow
``controllable'' or ``assisting'', much like a probe, in the sense
that we can control the measurement operators $M_j^E$ and have access
to the measurement outcomes $j$. This is the case, for example, of the
previously mentioned quantum eraser and partial teleportation. The
corrected channel then can take the form
$\sum_j\mathcal{C}_j^\rho\circ\mE_j$, where $\mE_j(\rho)
=\Tr_E[U(\rho\otimes|0\>\<0|_E)U^\dag\ (\openone\otimes M_j^E)]$.  The
probability of getting the $j$-th outcome is equal to
$p(j)=\Tr[\mE_j(\rho)]$, the conditional output state is $\sigma_j
:=\mE_j(\rho)/p(j)$, and both depend explicitly on the input system
state $\rho$ and on the POVM $\mM=\{M_j^E\}_j$. In practical
situations the hypothesis of complete control on the environment is
clearly too strong: in all these cases we can separate the environment
system $E$ into a controllable probe $P$ and ``the rest'' $R$. Then
the POVM operators $M_j^E$ will have the form
$M_j^E=M_j^P\otimes\openone_R$. Notice however that the assisted
scenario, as motivated in~\cite{GW}, when assuming for granted that
the system interacts with a completely controllable environment only,
mainly focuses on the \emph{in principle} information dynamics
involved in the overall process, and not on the practical feasibility
of the correction scheme itself. Hence in the following, when we will
need the ``complete controllability hypothesis'', we will explicitly
call for it: as a rule, we will speak of ``probe'' if \emph{complete}
control is possible, while the word ``environment'' will be left for
all other cases. Notice, moreover, that for a (finite) $d$-dimensional
system $\sH_S$, a $d^2$-dimensional probe suffices to realize whatever
quantum evolution.

\emph{Information retrieval and erasure.---} First of all, let us fix
some notation.  Given a channel $\mE$ acting on states of the input
system $S$, from Eq.~(\ref{eq:unitary-real}) we can always construct
the so-called \emph{complementary} channel $\aE$, defined as
$\aE(\rho)=\Tr_S[U(\rho\otimes|0\>\<0|_E)U^\dag]$. It describes the
output state of the environment given that the input state of the
system was $\rho$. By Stinespring theorem~\cite{stine}, such a
complementary channel is unique up to a partial
isometry~\cite{holevo}. Hence, we can consider $\aE$ as being the
canonical complementary channel. Moreover, given a channel $\mE$
acting on states, there exists a unique \emph{dual} channel $\dual\mE$
acting on observables $O$, defined by the trace relation
$\Tr[\mE(\rho)\ O]= \Tr[\rho\ \dual\mE(O)]$, for all $\rho$.  The
trace-preserving condition becomes a unit-preserving condition, i.~e.
$\dual\mE(\openone)=\openone$. We then have four channels: the direct
one, i.~e. $\mE$; the dual one, i.~e.  $\dual\mE$; the complementary
one, i.~e. $\aE$; and the complementary dual one, i.~e. $\dual\aE$.
If we send through the channel $\mE$ an ensemble of quantum states
$\{\rho_i\}_i$, such that $\Tr[\rho_i]=p(i)$, $\sum_i\rho_i=\rho$, and
$\Tr[\rho]=1$, at the environment output branch will arrive
$\{\aE(\rho_i)\}_i$. We then perform a measurement on them by using a
POVM $\mM=\{M_j^E\}_j$, thus obtaining a joint probability distribution
$p(i,j)= \Tr[\aE(\rho_i)\ M_j^E]= \Tr[\rho_i\ \dual\aE(M_j^E)]$.
In the following, we will consistently use the index $i$ for the input
ensemble and the index $j$ for the environment outcomes. We now invoke
the complete controllability hypothesis, and choose rank-one POVM
elements, i.~e. $M_j^E=|\phi_j\>\<\phi_j|_E:=\phi_j^E$. In fact, such
a choice is necessary and sufficient to rule out the possibility of a
\emph{classical post-processing} of data~\cite{clean}, which could
artificially reduce the information transmission. Since in the
following we will be interested in the measurement \emph{minimizing}
the information transmission, the choice of a rank-one probe
measurement is definitely the appropriate one. Correspondingly, the
channel $\mE$ gets decomposed into pure contractive maps
$\mE_j(\rho)=E_j\rho E_j^\dag$, or in other words, the rank-one
measurement $\mM$ refines the channel $\mE$ into a \emph{pure
  instrument}~\cite{davies}.

In order to quantify the amount of information about the input
ensemble $\{\rho_i\}$ that the measure of $\mM$ retrieves from the
probe, it is natural to compute the \emph{mutual information} from
$\vec p(i,j)$ as $\mutinf=H(\vec p(i))+H(\vec p(j))-H(\vec p(i,j))$,
where $H(\vec q(k))=-\sum_kq(k)\log q(k)$ is the Shannon entropy of
the probability distribution $\vec q(k)$. If $\mutinf$ is close to
zero, then $\vec p(i,j)\approx\vec p(i)\vec p(j)$, that is,
$p(i,j)\approx p(i)p(j)$ for all $i,j$, namely, the outcomes of the
measurement $\mM$ on the probe are almost independent of the input
ensemble $\{\rho_i\}_i$. It means that the information transmission is
poor, and if the same holds for all possible ensemble realizations
$\{\rho_i\}_i$ of $\rho$, we say that the measurement $\mM$ performs a
good \emph{erasure} with respect to the input state $\rho$. Here the
mutual information equals the \emph{relative entropy} $D(\vec
p(i,j)\|\vec p(i)\vec p(j))$ between the joint distribution $\vec
p(i,j)$ and the factorized one $\vec p(i)\vec p(j)$, where $D(\vec
r(k)\|\vec s(k))$ is defined for two probability distributions $\vec
r(k)$ and $\vec s(k)$ as $D(\vec r(k)\|\vec s(k))=\sum_kr(k)\log
r(k)/s(k)$,~\cite{cover}. If $s(k)=0$ for some $k$ for which $r(k)>0$,
then $D(\vec r(k)\|\vec s(k))$ diverges. In our case, however, this
will never be the case and the following inequalities will play a
central role~\cite{cover,hay,auden}
\begin{equation}\label{eq:ent-ineq}
  2^{-1}\N{\vec r(k)-\vec s(k)}^2_1\le D(\vec r(k)\|\vec
  s(k))\le\beta^{-1}\N{\vec r(k)-\vec s(k)}^2_1,
\end{equation}
where $\beta=\min_ks(k)>0$, and $\N{\vec r(k)-\vec
  s(k)}_1=\sum_k|r(k)-s(k)|$.


\emph{Correction efficiency.---} A useful quantity to judge the
capability of a channel $\mE$ in faithfully and coherently
transmitting an input state $\rho$, is given by the \emph{entanglement
  fidelity}~\cite{chanfid} $F_e(\rho)$ defined as
$F_e(\rho)=\Tr[|\Omega\>\<\Omega|\
(\mE\otimes\mathcal{I})(|\Omega\>\<\Omega|)]$, where $\mathcal{I}$ is
the identity (ideal) channel and $|\Omega\>$ is a purification of
$\rho$. If $F_e(\rho)$ is close to one, then the channel $\mE$ acts
quite like the identity channel on the support of
$\rho$,~\cite{chanfid}. Starting from a Kraus decomposition as in
Eq.~(\ref{eq:kraus}), with few calculations we find that
$F_e(\rho)=\sum_k|\Tr \rho E_k|^2$, and since it does not depend on
the particular decomposition $\{E_k\}$ chosen, it is an
\emph{intrinsic} property of the channel. The following simple upper
bound then comes from an application of a Cauchy-Schwartz--type
inequality
\begin{equation}\label{eq:fcorr}
  F_e(\rho)\le\sum_k(\Tr|\rho E_k|)^2:=F_{e,\textrm{a}}(\rho)\le 1,
\end{equation}
where $|\rho E_k|$ is the positive part of the polar decomposition
$\rho E_k=U_k^\rho|\rho E_k|$, for unitary $U_k^\rho$. With an
environment-assisted correction scheme, it is indeed possible to reach
such an upper bound, that we therefore call $F_{e,\textrm{a}}(\rho)$:
we have to measure on the probe the rank-one POVM corresponding to the
Kraus decomposition $\{E_j\}_j$, and to choose the conditional
correcting channels $\mathcal{C}_j^\rho$ to be equal
to the unitary channels
$\mathcal{C}_j^\rho(\sigma)=(U_j^\rho){}^\dag\sigma U_j^\rho$, where
$U_j^\rho$ is the unitary part of the polar decomposition $\rho
E_j=U_j^\rho|\rho E_j|$. If $F_{e,\textrm{a}}(\rho)$ is close to one,
it means that the corrected channel $\sum_k\mathcal{C}_k^\rho(E_k\rho
E_k^\dag)$ acts much like the ideal channel on the support of
$\rho$. The tricky point now is that $F_{e,\textrm{a}}(\rho)$
\emph{does depend} on the particular Kraus decomposition $\{E_j\}_j$,
that is, on the measurement $\mM=\{\phi_j^E\}_j$ performed upon the
probe system. The natural question is which measurement $\mM$
maximizes $\Fcorr$. We will answer showing that $\Fcorr$ essentially
determines how well the measurement $\mM$ erases from the probe the
information about the input state $\rho$, and vice-versa. In other
words, $\Fcorr$ and the erasure efficiency are equivalent measures, in
the sense that the less information about the input state the
measurement $\mM$ collects, the higher the corresponding $\Fcorr$ is,
and vice-versa.

\emph{Main result.---} Let us now fix the input state $\rho$ with
ensemble realization $\{\rho_i\}_i$ and the probe POVM
$\mM=\{\phi_j^E\}_j$. Then
$K_j=\rho^{1/2}\dual\aE(\phi_j^E)\rho^{1/2}/p(j)$ turn out to be
normalized states, for all $j$, with $\sum_jp(j)K_j=\rho$. Moreover,
by noticing that $\dual\aE(\phi_j^E)=E_j^\dag E_j$ we can rewrite the
upper bound in Eq.~(\ref{eq:fcorr}) as $\Fcorr=\sum_jp(j)
\fid\left(\rho,K_j\right)^2$, where
$\fid(\rho,\sigma):=\Tr[\sqrt{\sqrt{\rho}\sigma\sqrt{\rho}}]$ is the
Uhlmann fidelity between two mixed states~\cite{uhlmann}. By
exploiting the well-known relation~\cite{hay} between the fidelity and
the trace norm of the difference, defined as
$\N{\rho-\sigma}_1=\Tr|\rho-\sigma|$, that is, $\fid(\rho,\sigma)^2
\le 1-2^{-2}\N{\rho-\sigma}_1^2$, together with
Eq.~(\ref{eq:ent-ineq}), we obtain the following chain of inequalities
\begin{equation}\label{eq:direct}
\begin{split}
  \Fcorr&\le 1-2^{-2}\sum_jp(j)\N{\rho-K_j}_1^2\\
  &\le 1-2^{-2}\sum_{j}p(j)\left(\sum_i|p(i)-p(i|j)|\right)^2\\
  &\le 1-2^{-2}\beta\sum_jp(j)D(\vec p(i|j)\|\vec p(i))\\
  &=1-2^{-2}\beta\mutinf\le 1,
\end{split}
\end{equation}
where $\beta=\min_ip(i)>0$. In the second inequality we used the fact
that the trace distance between two states is never smaller than the
trace distance between the probability distributions obtained by
measuring the same POVM $\{\rho^{-1/2}\rho_i\rho^{-1/2}\}_i$ on both
states; notice that $\{\rho^{-1/2}\rho_i\rho^{-1/2}\}_i$ is a
well-defined POVM on the support of $\rho$, since
$\rho^{-1/2}\rho_i\rho^{-1/2} \ge 0$ and $\sum_i\rho^{-1/2}\rho_i
\rho^{-1/2}=\left.\openone\right|_{\supp(\rho)}$,~\cite{hjw}. In the
last equality we used the trivial identity $\sum_jp(j)D(\vec
p(i|j)\|\vec p(i))=D(\vec p(i,j)\|\vec p(i)\vec p(j))$. Notice
moreover that inequality~(\ref{eq:direct}) holds for every ensemble
realization $\{\rho_i\}_i$ of $\rho$.

Equation~(\ref{eq:direct}) informs us that if $\Fcorr$ is sufficiently
close to one (but it can be also \emph{strictly less} than one) for a
particular probe POVM $\mM$, then the corresponding information
transmission from the system to the probe is close to zero for
every ensemble realization $\{\rho_i\}_i$ of $\rho$, that is, the
measurement $\mM=\{\phi_j^E\}_j$ is \emph{erasing} the information about
the input state $\rho$ registered into the probe during the
interaction. Equivalently, non-null information extraction always
causes disturbance on the input ensemble, even allowing
input-dependent environment-assisted correction schemes. Our approach
hence embodies a quite general \emph{information-disturbance
  tradeoff}. It is worth stressing here that even if we drop the
complete controllability hypothesis, the same conclusions are true,
because the information extracted by the ``uncontrolled'' POVM
$\{\phi_j^P\otimes \openone_R\}_j$ is clearly less than or equal to
the information extracted by a fully controlled POVM $\{\phi^E_j\}_j$:
averaging decreases information.

The converse argument runs as follows. We have to check that, given
the channel and its realization as an interaction with a probe, for a
suitable probe rank-one POVM $\mM=\{\phi_j^E\}_j$, the information
transmission is poor \emph{for all possible} ensemble realizations
$\{\rho_i\}_i$ of a given input $\rho$. Luckily enough, we can
restrict our attention to just one particular set $\{\rho_i\}_i$ of
input states, with $\sum_i\rho_i=\rho$, being also
\emph{informationally complete}, that is, such that every operator $O$
on the support of $\rho$ is uniquely defined by its expectation values
on such ensemble. The existence of this kind of ensembles for every
finite dimension has been constructively proved in Ref.~\cite{ic}. The
reconstruction formula then holds
\begin{equation}\label{eq:recon}
  O=\sum_i\Tr[O\ \rho^{-1/2}\rho_i\rho^{-1/2}]\rho'_i,
\end{equation}
where the operators $\{\rho'_i\}_i$ are limited and hermitian but
neither positive definite nor semi-definite, in general. By
exploiting the well-known inequality $\fid(\rho,\sigma)^2 \ge
1-\N{\rho-\sigma}_1$,~\cite{hay}, we have the following relations
\begin{equation}\label{eq:converse}
\begin{split}
  \Fcorr&\ge 1-\sum_jp(j)\N{\rho-K_j}_1\\
  &=1-\sum_{ij}p(j)\N{p(i)\rho'_i-p(i|j)\rho'_i}_1\\
  &\ge 1-|\Gamma|\sum_{ij}p(j)|p(i)-p(i|j)|\\
  &\ge 1-\sqrt{2}|\Gamma|\sqrt{\mutinf},
\end{split}
\end{equation}
where $|\Gamma|=\max_i\N{\rho'_i}_1<\infty$. In the first equality we
used the reconstruction formula~(\ref{eq:recon}). Notice that, if
$\mutinf$ is sufficiently close to zero for one particular
informationally complete input ensemble realization $\{\rho_i\}_i$, then
$\Fcorr$ is close to one, and, by Eq.~(\ref{eq:direct}), $\mutinf$ is
also close to zero for all possible input ensemble realizations of
$\rho$. The implications then turn out to be
equivalences~\cite{nota4}.

We can hence conclude, by stating that \emph{for every channel
  realized as an interaction of an input state $\rho$ with a probe,
  even if perfect quantum erasure is impossible, the more a given POVM
  erases from the probe the classical information which can be carried
  by the input state and get stored in the probe during the
  interaction, the closer (on the support of $\rho$) the corresponding
  corrected channel is with respect to the ideal one, and
  vice-versa}. To find the optimal erasure measurement for a given
channel and a given input state remains an open problem. Incidentally,
it is worth noticing that if we consider $\rho=\openone/d$---indeed
just an invertible $\rho$ suffices---, then the quantum capacity of
the corresponding corrected channel is maximum over the whole input
Hilbert space $\sH_S$.
It is possible to achieve
\emph{perfect} erasure, that is
$F_{e,\textrm{a}}^\mathbf{M}(\openone/d)=1$, if and only if the
channel admits a random-unitary decomposition~\cite{GW}, that is,
$\mE(\rho)=\sum_jp(j)U_j\rho U_j^\dag$, for some probability
distribution $\vec p(j)$ and unitary operators $\{U_j\}_j$. In this
case, for the corresponding measurement $\mM$, $\mutinf$ is rigorously
zero \emph{for every} possible input ensemble~\cite{pla}.

\emph{Acknowledgments.---} Discussions with M~Hayashi, L~Maccone, and
M~F~Sacchi are gratefully acknowledged. The author also thanks the
anonymous Referees whose comments greatly contributed in clarifying
the manuscript. Japan Science and Technology Agency is acknowledged
for support through the ERATO-SORST Quantum Computation and
Information Project.

\appendix


\begin{thebibliography}{99}
\bibitem{nota4} Formally speaking, the two quantities $(1-\Fcorr)$ and
  $\mathscr I(\rho,\mM)
  :=\max_{\{\rho_i\}_i:\sum_i\rho_i=\rho}\mutinf$ are equivalent,
  since $a(\rho) \mathscr I(\rho,\mM)^2\le (1-\Fcorr)^2 \le
  b(\rho)\mathscr I(\rho,\mM)$, with $0<a(\rho)<b(\rho)<\infty$.
\end{thebibliography}
\end{document}